\documentclass[letter]{emulateapj} 

\usepackage{relsize}

\begin{document}


\title{Instantaneous starburst of the massive clusters Westerlund\,1 and NGC\,3603\,YC}


\author{Natalia Kudryavtseva\altaffilmark{1,*}, Wolfgang Brandner\altaffilmark{1}, Mario Gennaro\altaffilmark{1}, Boyke Rochau\altaffilmark{1}, Andrea Stolte\altaffilmark{2}, Morten Andersen\altaffilmark{3}, Nicola Da Rio\altaffilmark{3}, Thomas Henning\altaffilmark{1}, Emanuele Tognelli\altaffilmark{4}, David Hogg\altaffilmark{5}, Simon Clark\altaffilmark{6},and Rens Waters\altaffilmark{7}}

\altaffiltext{1}{Max-Planck-Institut f\"ur Astronomie, K\"onigstuhl 17, 69117 Heidelberg, Germany}
\altaffiltext{2}{Argelander Institut f\"ur Astronomie, Auf dem H\"ugel 71, 53121 Bonn, Germany}
\altaffiltext{3}{European Space Agency, Space Science Department, Keplerlaan 1, 2200 AG Noordwijk, The Netherlands}
\altaffiltext{4}{Department of Physics, University of Pisa, Largo Bruno Pontecorvo 3, 56127, Pisa, Italy}
\altaffiltext{5}{Center for Cosmology and Particle Physics, Department of Physics, New York University, 4 Washington Place, Room 424, New York, NY 10003, USA}
\altaffiltext{6}{Department of Physics and Astronomy, The Open University, Milton Keynes, MK7 6AA, UK}
\altaffiltext{7}{Sterrenkundig Instituut Anton Pannekoek, University of Amsterdam, Science Park 904, 1098 Amsterdam, The Netherlands} 
\altaffiltext{*}{Member of the International Max Planck Research School for Astronomy and Cosmic Physics at the University of Heidelberg.} 





\begin{abstract}
We present a new method to determine the age spread of resolved stellar populations in a starburst cluster. The method relies on a two-step process. In the first step, kinematic members of the cluster are identified based on multi-epoch astrometric monitoring. In the second step, a Bayesian analysis is carried out, comparing the observed photometric sequence of cluster members with sets of theoretical isochrones. When applying this methodology to optical and near-infrared high angular resolution \textit{Hubble Space Telescope (HST)} and adaptive optics observations of the $\sim$5\,Myr old starburst cluster Westerlund\,1 and $\sim$2\,Myr old starburst cluster NGC\,3603\,YC, we derive upper limits for the age spreads of 0.4 and 0.1\,Myr, respectively. The results strongly suggest that star formation in these starburst clusters happened almost instantaneously.
\end{abstract}





\section{Introduction}

Our understanding of star formation has progressed considerably in recent years (e.g., \citealt{2011ApJ...732...20K,2011ApJ...742L...9C}), though the exact sequence of the formation of individual stars during a star formation event is not yet entirely understood. In particular, it is still unknown if low-mass stars tend to form later than high-mass stars (e.g., \citealt{2001ApJ...550L..77K}), or if high-mass stars form last, resulting in a rapid termination of star formation (e.g., \citealt{2007ARA&A..45..481Z}).

The age spread of a cluster's stellar population is a good indicator of the overall duration of the star formation process in the cluster. Studies of star-forming regions have reported different results: from a single age, as in NGC 4103 \citep{1996JRASC..90Q.329F} to age spreads of 2-4\,Myr, e.g., in LH95 \citep{2010ApJ...723..166D}, the Orion Nebula Cluster \citep{2011A&A...534A..83R}, W3 Main \citep{2012ApJ...744...87B}, and even larger age spreads of tens of Myr, like in the Pleiades star cluster \citep{1998A&A...332..575B}. The broad range of age spreads might indicate the existence of different star formation scenarios for different star-forming environments. At the same time, discrepant results on the age spread in individual regions indicate that both observational and theoretical (e.g., \citealt{2009MNRAS.399..432N}) biases and the overall methodology used to assign ages to individual stars, might also be of importance. Observational difficulties include small number statistics, contamination of samples by field stars, variable extinction and intrinsic infrared excess, unresolved binaries and insufficient characterization of photometric uncertainties originating in varying degrees of crowding. 

For our analysis, we have selected Westerlund\,1 (Wd\,1) and NGC 3603 YC, two of the most populous and massive Galactic starburst clusters (e.g., \citealt{2005A&A...434..949C,2008AJ....135..878M}). The clusters have half-mass radii of $\approx$1 and 0.5\,pc, respectively. They are composed of more than 10,000 stars each, a large fraction of which can be resolved individually via ground-based adaptive optics and \textit{Hubble Space Telescope (HST)} observations from space. It is still unclear how such compact clusters have been formed and on which time-scales. For NGC 3603 YC, \citet{2004AJ....128..765S} suggest a single age of 1\,Myr from isochrone fitting to the pre-main sequence (PMS) transition region and a single burst of star formation, while \citet{2010ApJ...720.1108B} report a 10\,Myr age spread for the PMS population, and therefore two distinct episodes of star formation. For Wd\,1 recent studies give an age in the range 3-6\,Myr \citep{2008A&A...478..137B, 2011MNRAS.412.2469G, 2010A&A...516A..78N}, and an age spread of less than 1\,Myr \citep{2010A&A...516A..78N}. The latter is based on spectral classification of Wd\,1's OB supergiant population.

\section{Method}
\label{sec:method}

The first step of our method is a proper-motion selection of cluster members on the basis of multi-epoch astrometric observations (e.g., \citealt{2001ApJ...560L..75B}). This enabled us to reject the majority of the contaminating field stars. Next, we apply Bayesian analysis to the photometry of cluster members with respect to theoretical isochrones to determine the probability distribution for the age of each member star, given its photometric properties (e.g., \citealt{2010ApJ...723..166D}). We modified the Bayesian method of \citet{2005A&A...436..127J} both by taking into account the cluster membership probability and by adjusting the mass function (MF) by a completeness factor. The posterior probability of the $i$-th star with magnitudes $J_i, K_{\mathrm{S}_i}$ to belong to the isochrone of age $t$ is

\begin{eqnarray}
p( t | J_i, K_{\mathrm{S}_i})=\int\limits_{box} p(J_i, K_{\mathrm{S}_i} | M, t) \cdot  \xi(M | \alpha, t) dM,
\end{eqnarray}
where $M$ is the initial stellar mass and $\xi(M | \alpha, t)$ is the stellar MF with a slope $\alpha$. The integration region is a "box", i.e., a rectangular area in color-magnitude (CM) space: $K_{\mathrm{S}_{min}}\,<\,K_\mathrm{S}\,<\,K_{\mathrm{S}_{max}}, (J-K_\mathrm{S})_{min}\,<\,(J-K_\mathrm{S})\,<\,(J-K_\mathrm{S})_{max}$. The limits of this area define a mass range which is considered during integration. This box has been chosen in order to exclude field star contaminants based on their colors. To account for residual field star contamination, the first multiplier inside the integral is defined as:
\begin{eqnarray} \label{eq:P(J,K|M,t)}
p(J_i, K_{\mathrm{S}_i}| M, t) = P_{good} \cdot  p(J_i, K_{\mathrm{S}_i}| M, t, cluster)\nonumber \\
+ (1-P_{good}) \cdot  p(J_i, K_{\mathrm{S}_i} | field),
\end{eqnarray}
where $P_{good}$ is the probability of a star to be a cluster member and $(1-P_{good})$ is the probability to be a field star.\newline
From the normalization conditions 
\begin{eqnarray} 
\int\limits_{box} p(J_i, K_{\mathrm{S}_i}| M, t, cluster) dJ dK_\mathrm{S} = 1,  
\end{eqnarray}
\begin{eqnarray}
\int\limits_{box} p(J_i, K_{\mathrm{S}_i}| field) dJ dK_\mathrm{S} = 1 
\end{eqnarray}
we derive the probability that a cluster member or field star is found in a given position on the color-magnitude diagram (CMD) as
\begin{eqnarray}\label{eq:P(cluster)}
p(J_i, K_{\mathrm{S}_i}| M, t, cluster) = \frac{1}{ 2\pi\sigma_{J_i}\sigma_{K_{\mathrm{S}_i}} }\times \nonumber \\
e^{-\frac{1}{2}\big[\big(\frac{J(M,t)-J_i}{\sigma_{J_i}}\big)^2+\big(\frac{K_\mathrm{S}(M,t)- K_{\mathrm{S}_i}}{\sigma_{K_{\mathrm{S}_i}}}\big)^2\big]}, 
\end{eqnarray}
\begin{eqnarray}
p(J_i,  K_{\mathrm{S}_i}| field) = \frac{1}{ \Delta(K_\mathrm{S}) \Delta(J-K_\mathrm{S})}, 
\end{eqnarray}
where $\sigma_{J_i},\sigma_{K_{\mathrm{S}_i}}$ are the photometric uncertainties of star\,$i$ and $J(M,t), K_\mathrm{S}(M,t)$ are the magnitudes of the theoretical isochrone, $\Delta(K_\mathrm{S})=K_{\mathrm{S}_{max}}-K_{\mathrm{S}_{min}}$, $\Delta(J-K_\mathrm{S})=(J-K_\mathrm{S})_{max}-(J-K_\mathrm{S})_{min}$.
The MF $\xi(M | \alpha, t)$ was modified by $compl(M | t)$ to include source incompleteness and has the following form:
\begin{eqnarray}\label{eq:mass_funct}
\xi(M| \alpha, t)= B \cdot M^{-\alpha}\cdot  compl(M | t), 
\end{eqnarray}
where B was derived from the normalization condition 
\begin{eqnarray}
\int\limits_{box} \xi(M| \alpha, t) dM = 1 
\end{eqnarray}
and $compl(M | t)$ from completeness simulations (see Section~\ref{sec:completeness}).

By multiplying the individual age distributions, we obtain the global probability function for the cluster's age $t$:\newline
\begin{eqnarray}\label{eq:L(t)}
L(t)=\displaystyle\prod_{i}p( t | J_i, K_{\mathrm{S}_i}).
\end{eqnarray}

\section{Observations and data reduction} 

\subsection{Observations of Westerlund\,1 and NGC\,3603\,YC}

Near-infrared adaptive optics observations of the central region of Wd\,1 were carried out in 2003 April, using NACO at the Very Large Telescope (VLT). $K_\mathrm{S}$ observations with a plate scale of 27 mas/pixel and a field of view (FOV) of $27"\times27"$ (corresponding to 0.5pc$\times$0.5pc at 4.0\,kpc distance; \citep{2011MNRAS.412.2469G}) were centered on R.A.$(2000) = 16^h47^m06^s.5$, decl.$(2000) = - 45^\circ 51' 00'' $. $K_\mathrm{S}$ frames with integration times of 1minute were co-added, resulting in a total integration time of 5 minutes.

In 2010 August (epoch difference 7.3\,yr), Wd\,1 was observed with the \textit{HST} Wide Field Camera\,3 (WFC3/IR) in the F125W band with a plate scale of 130 mas/pixel. The final image consists of seven individual exposures with small ($<$10\arcsec) offsets to compensate for bad pixels, with a total integration time of 2444\,s. The overlapping FOV with NACO is $\approx 18'' \times 24''$. For a detailed description of the full data set and data reduction, we refer to the work of M.Andersen et al. (2012, in preparation).

Positions and magnitudes of the stars were determined using DAOPHOT \citep{1987PASP...99..191S}. Instrumental magnitudes were calibrated against Two Micron All Sky Survey, using suitable stars identified on the NTT/SOFI data from the work by \cite{2011MNRAS.412.2469G} as secondary photometric standards.
To align the two data sets, identical bright stars with a high probability of being cluster members were identified. The geometric transformation of the NACO pixel coordinates to the WFC3/IR system was based on fitting third-order polynomials with cross-terms in \textit{x} and \textit{y}. For bright stars with $m_{K_\mathrm{S}}\le16.0$ \,mag this resulted in an rms of $\approx$0.036 WFC3/IR pixel ($\approx$4.6\,mas) for the astrometric offset between the WFC3/IR coordinates and the transformed NACO coordinates.

Images of NGC\,3603\,YC were taken with \textit{HST}'s Wide Field Planetary Camera 2 (WFPC2) with an image scale of 45.5\,mas/pixel. The first epoch observations in the filters F547M and F814W were separated by 10.15\,yr from the second epoch observations in F555W and F814W filters. The analysis was carried out for the core ($<$0.5\,pc) of NGC\,3603\,YC. The details of the data reduction process have already been described by \cite{2010ApJ...716L..90R}.

\subsection{Completeness Correction}
\label{sec:completeness}

As the matched \textit{HST}/VLT data set for Wd\,1 is limited by the detection sensitivity in the $J$ band (F125W), the completeness simulation was carried out only for the WFC3/IR data. Completeness simulations were done for the magnitude range $m_{J}$ from 14.0 to 23.0\,mag. For each simulation, 25 artificial stars were added at random positions in the image, the image was then analyzed using DAOPHOT, and the position and magnitude of the recovered stars recorded. This procedure was repeated 1600 times, i.e.\ encompassing 40,000 artificial stars in total. The recovery fractions were determined following the steps outlined by \cite{2011MNRAS.412.2469G}. We calculated the average completeness for the WFC3/IR area overlapping with the NACO frame as a function of $m_{J}$ and fitted this by a Fermi function $ compl(M | t)= \frac{A_0}{e^{\frac{m_{J}(M,t)-A_1}{A_2}}+1} $.

Completeness simulations for NGC\,3603\,YC were done for F555W. Ten artificial stars were added at each of 50 runs, in total 500 stars for each $m_{F555W}$ magnitude bin between 16.0 and 23.0\,mag. We calculated the average completeness for \textit{HST}/WFPC2 image as a function of $m_{F555W}$ and fitted this by a Fermi function. 

\subsection{Proper-motion Selection}

The main contaminants apparent in CMDs of the starburst clusters are dwarf stars in the foreground and giants in the background. Due to galactic rotation, their proper motion is different from that of cluster stars. Velocity dispersions of disk and halo stars are $\approx$50\,km/s to 150\,km/s \citep{2011MNRAS.412.1203N}.

\begin{figure}
\epsscale{0.8}
\plotone{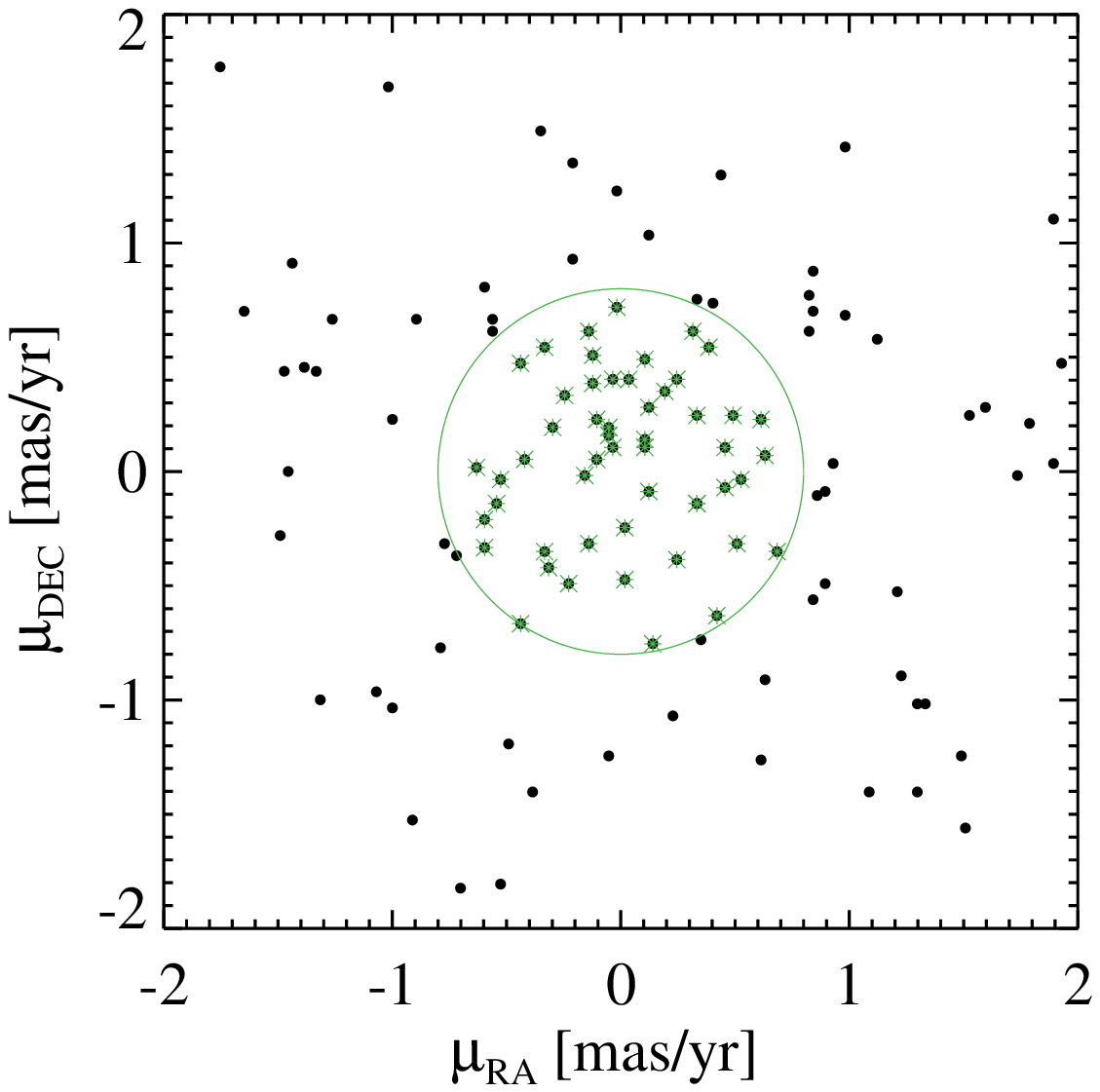}
\caption{Proper motion diagram for Wd\,1. Stars with proper motions $<$0.8 mas/year are marked by green asterisks.} \label{label:ppm}
\end{figure}

For Wd\,1 (see Figure~\ref{label:ppm}), our selection criterion for the astrometric residual of 5.8 mas corresponds to a proper motion of 0.8 mas/yr in the cluster rest frame (or 15.2 km/s at a distance of 4.0 kpc), which is higher than Wd\,1's internal velocity dispersion of 2.1 km/s \citep{2012A&A...539A...5C}. Such a selection provides an effective discriminant between cluster members and field stars (see Figure~\ref{label:fig_CMDs}a).

For NGC\,3603\,YC, we used the result of \cite{2010ApJ...716L..90R}, which is based on proper motions over an epoch difference of 10.15 years. The authors calculated cluster membership probabilities as described by \cite{1988AJ.....95.1755J}, and considered the stars with probabilities $>$90\% as cluster members. 

\section{CMD for Wd\,1 and NGC\,3603\,YC}

\begin{figure}
\epsscale{0.9}
\plotone{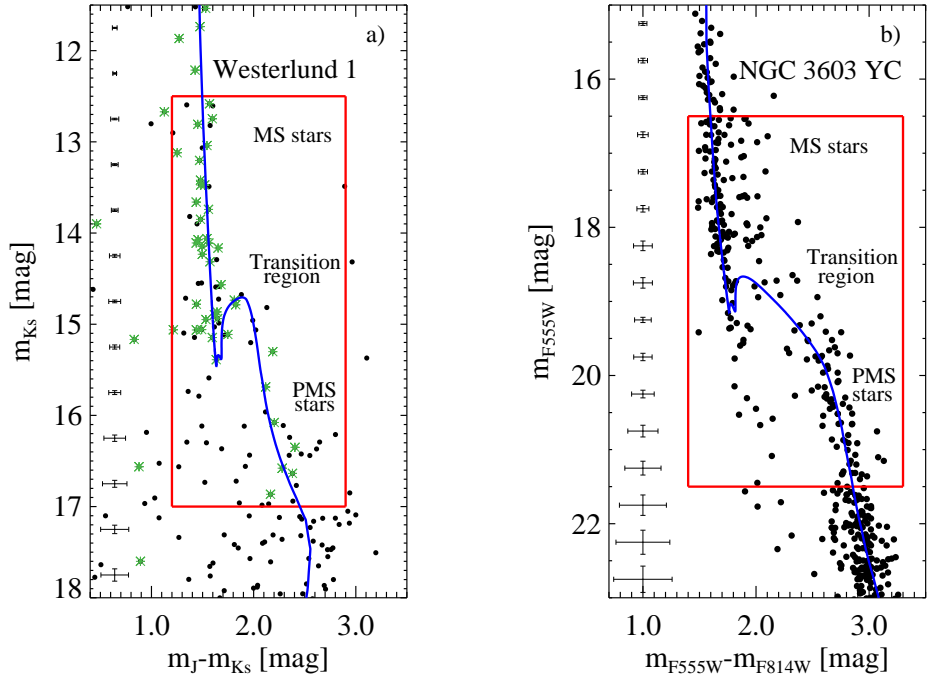}
\caption{Color-magnitude diagrams of Wd\,1 (a) and NGC\,3603\,YC (b). Error bars indicate typical errors in color and magnitude. Red boxes show the regions that were taken for the analysis of the age spread. Proper-motion-selected stars for Wd\,1 are marked by green asterisks. The CMD of NGC\,3603\,YC includes only stars with cluster membership probabilities $>$90\%. FRANEC-Padova isochrones (blue) are overplotted. For Wd\,1 this is a 5.0\,Myr isochrone at DM=13.0\,mag and for NGC\,3603\,YC a 2.0\,Myr isochrone at DM=14.1\,mag.} \label{label:fig_CMDs}
\end{figure}

The CMD for Wd\,1 is presented in Figure~\ref{label:fig_CMDs}a. For our further analysis we consider only the region with $12.5 < m_{K_\mathrm{S}} < 17.0$\,mag and $ 1.2 < m_{J}-m_{K_\mathrm{S}} < 2.9 $\,mag (red box in Figure~\ref{label:fig_CMDs}), which comprises 41 stars with masses in the range from 0.5 to 11.5 $M_{\sun}$. Brighter stars were excluded because of saturation, fainter stars because of lower signal-to-noise ratio and hence larger photometric and astrometric uncertainties. The sample includes main sequence (MS), Pre-MS, and transition region stars, which have terminated their fully-convective Hayashi phase and are rapidly moving towards the MS.
The CMD for NGC\,3603\,YC (Figure~\ref{label:fig_CMDs}b) is derived from second epoch observations in F555W and F814W. For the age spread determination we selected the region with $16.5 < m_{F555W} < 21.5$\,mag and $1.4 < m_{F555W}-m_{F814W} < 3.3$\,mag (red box), which comprises 228 stars with masses from 0.8 to 6.5 $M_{\sun}$.
Overplotted is the best fitting isochrone assuming a particular distance, for a solar metallicity $Z$=0.015, calculated from the latest version of FRANEC evolutionary models\footnote{FRANEC models in a wide range of masses and ages, for several chemical compositions and mixing length parameter, are available at \url{http://astro.df.unipi.it/stellar-models/}} \citep{2011A&A...533A.109T}, adopting a mixing length value of ML$=1.68$. We supplement the FRANEC models with Padova models \citep{2008A&A...482..883M} for masses $M>7M_{\sun}$. The FRANEC models have been transformed into the observational plane using spectra from ATLAS9 model atmospheres \citep{2004astro.ph..5087C}. For the analysis, we used isochrones with 0.1\,Myr spacing, covering an age range from 0.5 to 6\,Myr. 

\section{Age likelihood for Wd\,1 and NGC 3603 YC}

The application of our method (Section~\ref{sec:method}) to Wd\,1 and NGC 3603\,YC reveals a slight degeneracy between the cluster's distance and age. Since all cluster members should be at virtually the same distance, we can select a set of distances, and then analyze the resulting cluster age and age spread for each particular distance. 

In order to evaluate $P_{good}$ in Equation(\ref{eq:P(J,K|M,t)}) for Wd\,1 we estimated the density of stars in the CMD regions adjacent to the red box (Figure~\ref{label:fig_CMDs}, $ m_{J}-m_{K_\mathrm{S}}\,>\,2.9$\,mag and $m_{J}-m_{K_\mathrm{S}}\,<\,1.2 $\,mag), where the stars are apparently non-cluster members. The extinction value of $A_{K_\mathrm{S}}=1.1$\,mag \citep{2008A&A...478..137B} was assumed the same for all cluster members. The extinction law was taken from \cite{1985ApJ...288..618R}. For the MF slope in Equation(\ref{eq:mass_funct}) we assumed $\alpha=1.42$ (where Salpeter slope would correspond to 2.3), as derived from near-infrared adaptive optics observations of the central region of Wd\,1 (N.Kudryavtseva et al., 2012, in preparation). In order to quantify the photometric error $\sigma$ in Equation(\ref{eq:P(cluster)}), we used the results of the artificial star experiments to compare the known input magnitudes with output magnitudes recovered by DAOPHOT. A more detailed explanation of the procedure is described in \cite{2011MNRAS.412.2469G}. The maximum photometric errors we got are 0.05\,mag in $K_\mathrm{S}$ and 0.18\,mag in $J$ for Wd1, and 0.17\,mag in both bands of NGC\,3603\,YC. 

\begin{figure}
\epsscale{0.9}
\plotone{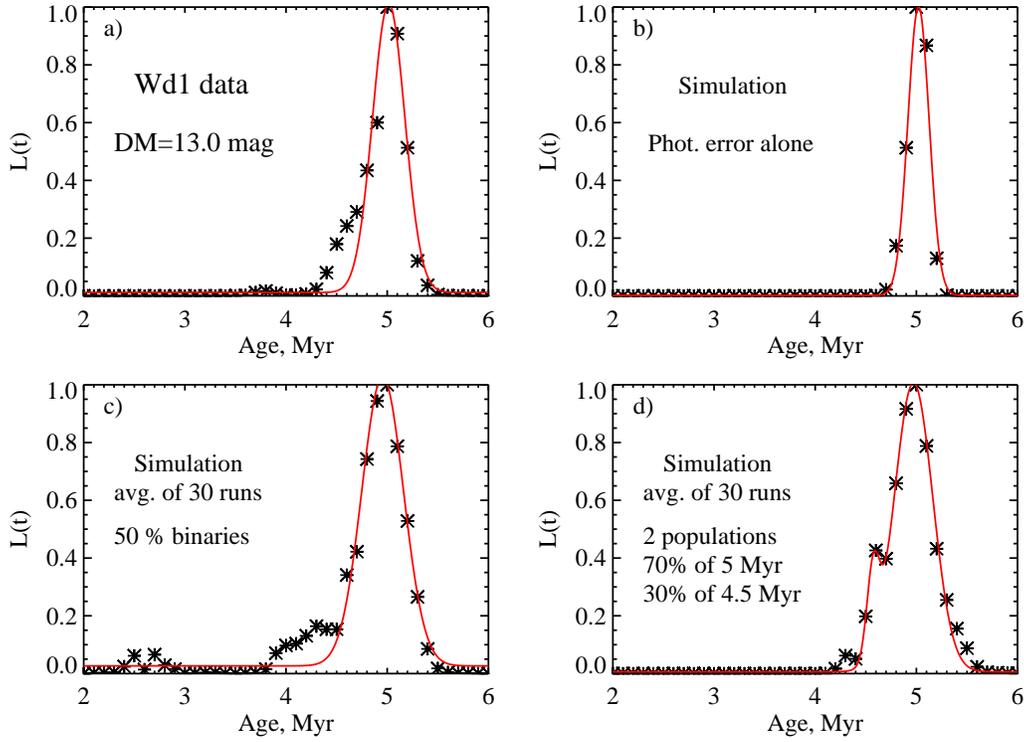}
\caption{ (a) Normalized $L(t)$ for Wd\,1 at DM=13.0\,mag. The most probable age is 5.0\,Myr. The red curve is a fitted Gaussian.
(b) Normalized $L(t)$ for the simulated 5.0\,Myr population at DM=13.0\,mag.
(c) Normalized $L(t)$ for the simulated 5.0\,Myr population at DM=13.0\,mag with a binarity fraction of 50\%.  
(d) Normalized $L(t)$ for the simulated population with 70\% of the stars being 5.0\,Myr old and 30\% being 4.5\,Myr old.} \label{label:6cuts}
\end{figure}

The $L(t)$ function we derived from Equation(\ref{eq:L(t)}) for Wd\,1 at a distance modulus (DM) of 13.0\,mag (4.0\,kpc) is presented in Figure~\ref{label:6cuts}a. The full width at half maximum (FWHM) of a Gaussian fitted to $L(t)$ (red line in Figure~\ref{label:6cuts}a) is 0.4\,Myr. For $12.8\leq$ DM $\leq13.2$mag we got similar result.

\begin{figure}
\epsscale{0.55}
\plotone{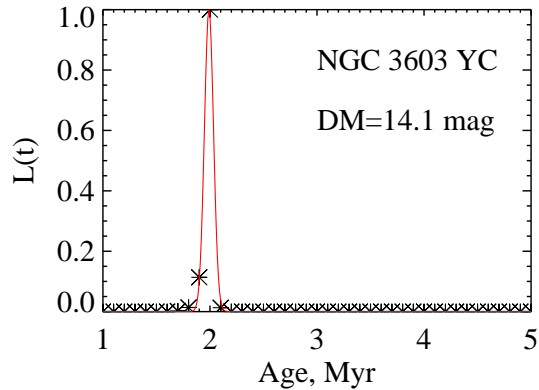}
\caption{Normalized $L(t)$ for  NGC\,3603\,YC at DM=14.1\,mag. The most probable age is 2.0\,Myr. The red curve is a fitted Gaussian function.} \label{label:3cuts ngc}
\end{figure}

The $L(t)$ function for NGC\,3603\,YC at DM=14.1\,mag is presented in Figure~\ref{label:3cuts ngc}. As contamination by field stars was already significantly reduced by proper motion selection, $P_{good}$ in Equation(\ref{eq:P(J,K|M,t)}) was estimated, assuming that cluster members are the stars with cluster membership probabilities $>$98\%. The extinction $A_v=4.9$\,mag is in agreement with the results by \cite{2004AJ....127.1014S} and \cite{2010ApJ...716L..90R} and the relative extinction relations from \cite{1998ApJ...500..525S}. We used the same $\alpha=1.9$ as in \cite{2006AJ....132..253S} for the MF slope. The maximum FWHM we derive for NGC 3603\,YC at $13.9\leq$ DM $\leq14.3$mag is 0.1\,Myr.\newline

\section{Broadening of the age likelihood function}

In the ideal case of a coeval population, which lies along an isochrone, and has no photometric errors, $L(t)$ from Equation(\ref{eq:L(t)}) would be a Dirac delta function. A number of observational and physical effects are potentially responsible for the $L(t)$ broadening. In this section, we model these effects in order to estimate the true age spread. \newline

\subsection{Photometric Error} 
\label{sec:phot.error}

 In order to quantify the broadening of $L(t)$ due to solely photometric uncertainties we first generated a number of cluster stars along an isochrone of a certain age in the CM space, and added random photometric errors. Random field stars were added to the data set with the same density as derived for real data. For this simulated data set, we applied the likelihood technique as described in Section~\ref{sec:method} and got an artificial $L(t)$ function (see Figure~\ref{label:6cuts}b). Only due to photometric errors our simulation on 5.0\,Myr population gave an $L(t)$ broadening in terms of FWHM equal to 0.25 Myr, hence the true age spread for Wd\,1 should be even less than 0.4\,Myr.

\subsection{Unresolved Binarity} 

There is considerable observational evidence that binary stars might constitute a significant fraction of a cluster population (e.g., \citealt{2008AJ....135.1934S}). As an unresolved binary combines the light of two stars, a binary will result in an offset in brightness by up to $-0.75$\,mag in the CMD compared to a single star. For non-equal mass systems, there might also be an offset in color. Both cases lead to a broadening of the observed cluster sequence on the actual CMD.

To test how the shape of $L(t)$ is affected by unresolved binarity, we generated stars in the CM space in a similar way as described above for photometric errors, but with a 0.75\,mag shift along the ordinate for 50\% of the artificial points. We choose 50\% as the mean value for the binarity fraction, as recent observations at least for massive population of Wd\,1 revealed a high rate (more than 40\%) of binary stars \citep{2009A&A...507.1585R}. We applied the same Bayesian analysis to the simulated stars as to the real data to determine $L(t)$. 

The normalized likelihood $L(t)$ we obtained after combining the results of 30 simulations on binarity for a 5.0\,Myr population at DM 13.0\,mag is shown in Figure~\ref{label:6cuts}c. It is clearly seen that binarity affects the shape of $L(t)$ by adding a pronounced wing to the left from the main peak. Hence, a small shoulder toward younger ages from the $L(t)$ maximum for Wd\,1 (Figure~\ref{label:6cuts}a) could be caused by unresolved binarity.

\subsection{Ongoing Accretion}

For young stellar populations with ages $\lesssim$10\,Myr, ongoing accretion and the accretion history of a star is also of importance. As discussed in, e.g.,  \cite{2011ApJ...738..140H}, low-mass objects, which gain mass through ongoing accretion and a non-accreting PMS star of the same mass, arrive at different positions in a Hertzsprung-Russell diagram. 

For starburst clusters like Wd\,1 and NGC\,3603\,YC, this effect seems to be strongly attenuated due to the presence of very luminous and massive O-type stars in the clusters. The fast winds and the ionizing radiation from these stars evaporate and remove circumstellar material around the low-mass cluster members, and clean out any remnant molecular gas in the cluster environment on short timescales of a few 10$^5$yr (e.g., \citealt{2004ApJ...611..360A,1998ApJ...499..758J}). This results in very little differential extinction across the starburst cluster, enabling us to constrain stellar properties using broadband photometry (\citealt{2004AJ....128..765S}). 

\subsection{Sensitivity to Age Spread}
\label{sec:2 populations}

In order to test the ability of our method to detect the real age spreads, we simulated a mixed cluster population with two different ages and assuming random photometric errors. We repeated this procedure 30 times and calculated the averaged normalized likelihood $L(t)$. The result for a mixed population with 70\% stars of age 5.0\,Myr and 30\% stars of age 4.5\,Myr is shown in Figure~\ref{label:6cuts}d. Hence, a bimodal population with an age difference $\ge$0.5\,Myr would manifest itself in a prominent secondary peak, but the latter is not revealed in case of Wd\,1 (see Figure~\ref{label:6cuts}a).

\section{Summary}

Our analysis highlights the importance of a good membership selection, background rejection and characterization when analyzing crowded field data. We emphasize that the photometric component of our analysis only works in young, massive clusters and older clusters with little differential extinction and absence of ongoing accretion. Other environments require detailed spectroscopic analyses to establish precise effective temperatures and luminosities for individual stars, as has been exemplified in the case of ONC \citep{1997AJ....113.1733H} and W3\,Main \citep{2012ApJ...744...87B}.

Isochrones based on FRANEC and Padova evolutionary models, and ATLAS9 atmospheric models provide a good match to observed cluster sequences in the age range from 1 to 5\,Myr and mass range from 0.6 to 14\,$M_\odot$ for solar metallicities. Extending the sample of cluster members to lower masses could help to benchmark evolutionary tracks and atmospheric models for young, low-mass stars and quite possibly even brown dwarfs.

Our analysis of the photometric sequence of cluster members yields an age spread of less than 0.4\,Myr for the $\approx$5.0\,Myr old Wd\,1 and less than 0.1\,Myr for the $\approx$2.0\,Myr old NGC\,3603\,YC. This is a strong indication that in both cases the clusters formed in a single event once a sufficient gas mass had been aggregated and compressed to overcome internal thermal, turbulent or magnetic support, and to initiate an avalanche-like star formation event. This finding seems to be in agreement with theoretical predictions that clusters with masses between $10^4$ and $10^5 M_{\sun}$ loose their residual gas on a timescale shorter than their crossing times \citep{2008MNRAS.384.1231B}. For Wd\,1 and NGC\,3603\,YC crossing times have been estimated to be of the order of 0.3\,Myr \citep{2008A&A...478..137B} and 0.03\,Myr \citep{2010IAUS..266...24P}, respectively. 




\clearpage



\end{document}